\documentclass{emulateapj} % for paper style
\usepackage{apjfonts}
\usepackage{epsfig,graphicx}

\newcommand{\sci}{{\small I}\ }

\slugcomment{Received ........; Accepted .........}

\shorttitle{ Opposite polarity in simulated sunspot penumbra}

\shortauthors{Bharti et al.}

\begin{document}

\title{ Opposite polarity magnetic field and convective downflows in a simulated sunspot penumbra}

\author{Lokesh Bharti$^{1}$ and Matthias Rempel$^{2}$ }
\affil{1. Bal Shiksha Sadan Society, 3, Ratakhet, Udaipur, India}
\affil{2. High Altitude Observatory, NCAR, P.O. Box 3000,
       Boulder, CO 80307, USA}
       \email{lokesh$\_$bharti@yahoo.co.in}

\begin{abstract}
Recent numerical simulations and observations of sunspots show a significant amount of
opposite polarity magnetic field within the sunspot penumbra. Most of the opposite polarity field is
associated with convective downflows. We present an analysis of 3D MHD simulations
through forward modeling of synthetic Stokes profiles of the Fe\sci 6301.5 \AA~ and Fe\sci 6302.5 \AA~
lines). The synthetic Stokes profiles are spatially and spectrally degraded
considering typical instrument properties. Line bisector shifts of the Fe\sci 6301.5 \AA~ line are used to determine line-of-sight velocities. Far wing magnetograms are constructed
from the Stokes V profiles of the Fe\sci 6302.5 \AA~ line. While we find an overall good agreement
between observations and simulations, the fraction of opposite polarity magnetic field, the downflow filling factor
and the opposite polarity-downflow association are strongly affected by spatial smearing and presence of strong gradients in the line-of-sight magnetic field and velocity. A significant fraction of opposite polarity magnetic field and downflows are hidden in the observations due to typical instrumental noise. Comparing simulations that differ
by more than a factor of two in grid spacing we find that these quantities are robust within the simulations.

\end{abstract}


\section{Introduction}

Over the past decade sunspot simulations have advanced significantly and reached a degree of sophistication at which details of
the observed fine structure in umbra and penumbra are reproduced (Rempel et al. 2009, Rempel 2011a, 2012).
These simulations suggest that the filamentation in the sunspot penumbra is a result of overturning convection
in the presence of a strongly inclined magnetic field. In these simulations overturning convection is the mechanism
that explains the brightness of the penumbra.

Bharti et al. (2011) used the simulations of Rempel (2011b) in order to
investigate if convective downflows in the sunspot penumbra can be detected with presently available instruments.
Their predictions were verified through observations of convective downflows by Joshi et al. (2011) and Scharmer et al. (2011).
In addition, recently discovered twisting motions in penumbral filaments (Ichimoto et al.
 2007) have been considered as indirect evidence of convection in the penumbral filaments (Zakharov et al. 2008, Bharti et al.
 2010). Using penumbral simulations in slab geometry (Rempel 2009), Bharti et
 al. 2012 showed that twisting motions are also present in simulated penumbrae. The advantage of such comparison is that we can understand details of such dynamical phenomena that are hidden from direct observations. Using this approach Bharti et al. (2012)
 found that the "twisting'' motions in penumbral filaments are caused by transverse oscillations that cause a sideways swaying of the filaments.

The simulations of Rempel (2012) showed that there is a significant amount
of opposite polarity magnetic field present in a simulated sunspot penumbra. This is caused by two effects:
submergence of field lines by strong convective downflows at the sides and submergence of field lines at the tail of filaments.
The presence of opposite polarity fields at the sides of penumbral filaments and their relation to convective
downflows have been confirmed in ground based and space based observations
(Scharmer et al. 2013, Ruiz Cobo \& Asensio Ramos 2013, Tiwari et al. 2013, Esteban et al. 2015). Such downflows
and magnetic fields are consistent with magnetoconvective models of the penumbra (Rempel 2009, 2012).
However, Franz and Schlichenmaier (2009, 2013) found most of the opposite polarity magnetic field
and downflows only in the outer part of penumbral filaments. Prior to above mentioned work Westerndrop Plaza et al. (1997, 2001),
del Toro Iniesta et al. (2001) and Langhans et al. (2005) reported on opposite polarity magnetic fields in the penumbra. However, the amount of
opposite polarity magnetic flux in observations and a detailed comparison with predections from currently available models remain an open question (see reviews by Solanki (2003), Thomas \&
Weiss (2004, 2008), Borrero \& Ichimoto (2011), Rempel \& Schlichenmaier (2011), Tiwari (2017), Kubo (2018)).

In this paper we aim to find the amount of opposite polarity field and its association
with convective downflows that is present in the penumbra of a simulated sunspot. We investigate
how these quantities depend on the resolution of MHD simulations and how their detectability
depends on the resolution of observations as well as the method used.

\section{Simulations, line synthesis and convolution}
\label{sec:simul}

Our investigation is based on the numerical sunspot models described in Rempel (2012).
These are simulations of a single 10$^{22}$Mx sunspot in a computational domain
with a horizontal extent of $49.152\times49.152$~Mm$^2$ and a vertical extent of
6.144 Mm.  In order to obtain extended penumbrae in a horizontally periodic
setting, these simulations use a top boundary condition (about 700 km above the
photosphere), which artificially increases the inclination angle of the magnetic
field compared to a potential field extrapolation. We consider here two models
with horizontal/vertical grid spacings of 32/16 km and 12/8 km. These models
are non-grey versions of the grey models presented in Rempel (2012). The model
with 32/16 km grid spacing was restarted from the equivalent model in Rempel (2012)
and evolved with non-grey radiative transfer for an additional 26 minutes; the model
with 12/8 km grid spacing was started from a 16/12 km model and evolved for an
additional 15 minutes with higher resolution and non-grey radiative transfer.

The data from these simulations consist of 3D data cubes, which we use for the forward
modeling of spectral lines, as well as 2D slices on geometric height and constant
$\tau$ surfaces. The data on constant $\tau$ surfaces are extracted from the 3D data cubes based
on the $\tau$-scale for a vertical ray using the opacity bin that contains most of the continuum
contribution (the no-grey radiative transfer uses a total of 4 opacity bins). We extract physical quantities
on a $\tau=\tau_0$- surface by assuming a Gaussian contribution function centered on $\tau_0$ with a
FWHM of  $\tau_0/3$. In the following analysis we will use 2D data extracted on the $\tau=1$ surface directly in the simulation. These
data serve as a reference, which does not involve any radiative transfer beyond computing a vertical
$\tau$-scale.

For the analysis presented in this paper, we follow the approach of Bharti et al. (2011).
We use simulated sunspot snapshots with 32 km ($1280\times1280$ pixels) and 12 km ($4096\times4096$ pixels) horizontal
grid spacing. The SPINOR code (Berdyugina et al. 2003) with STOPRO (STOkes
PROfiles) was used to determine synthetic line profiles. The code calculates the Stokes
parameters for spectral lines in local thermodynamic equilibrium (LTE).
We first compute full Stokes profiles for the Fe\sci 6301.5 \AA~ and Fe\sci 6302.5 \AA~ lines at the full resolution of the
simulations (32 km and 12 km), in a second step, Stokes profiles are spatially and spectrally degraded according to instrumental properties. For comparison
with Hinode SP observations at 0.32\arcsec resolution we use a point-spread function that considers central obscuration
and spider (van Noort 2012) for spatial convolution and a Gaussian
with an FWHM of 21.54~m\AA~ for the convolution in wavelength. The polarimetric
noise of 10$^{-3}$ $I_c$ is added to the synthetic Stokes profiles ($I_c$ is the mean
continuum intensity of granulation found outside the sunspot). These profiles
are sampled according to Hinode wavelength points (112). The convolution
gives $\sim$8\% RMS continuum contrast in the granulation outside the sunspot.
For comparison with the 1 m and 1.5 m circular aperture telescopes
we use the approach of Bharti et al. (2011) considering a Gaussian and a Lorentzian (cf. Shelyag et al. 2004).
In order to emulate the observed straylight characteristics of SST/GREGOR we chose a contribution from the Lorentzian such that the resulting RMS continuum contrast for the granulation outside the sunspot is 12\% and 8\%, respectively. For spectral smearing we use a Gaussian with 45~m\AA~ FWHM.

The above mentioned scheme is also used for spatial smearing of the $\tau$ slices at Hinode (0.5 m), 1 m and 1.5 m resolution.

We determine the bisector of the FeI 6301.5 \AA~ line profile for all pixels of the resulting two dimensional map.
The bisector is determined by calculating the Doppler shifts at 10 (relative) intensity levels of the line in steps of
10\% (the relative intensity at the line core and at the continuum level considered to be 0$\%$ and 100$\%$, respectively).
For our analysis we use the bisector velocity map at 80$\%$ intensity level since the higher bisector levels approach the continuum
level and are prone to noise. The mean velocity of the darkest part of the umbra is considered as zero velocity reference and subtracted
from the bisector map.

Center-of-gravity (COG) method (Uitenbroek 2003) is used to retrieve line-of-sight magnetic field for FeI 6302.5 \AA~ line at native, Hinode (0.5 m), 1 m and 1.5 m resolution respectively.

The Stokes V profiles are used to infer magnetic field of opposite polarity (relative to the polarity of the spot).
We follow the approach of Ichimoto et al. (2007) and Franz \& Schlichenmaier (2013)
to construct FeI 6302.5 \AA~ wing magnetograms. The fraction of opposite polarity field in the penumbra for Hinode (0.5 m) at 30 m\AA~ and for 1 m and 1.5 m cases found to be maximum
 at 20 m\AA~ from the FeI 6302.5 \AA~ line center.

We do not consider any velocity signal -100 ms$^{-1}$ $\leq$ $V_{\rm LOS}$ $\leq$ 100 ms$^{-1}$ for the LOS velocity maps.
Stokes V signal above the 3$\sigma$ noise level being taken in to account to construct far wing magnetograms.

We also used the method used by Franz \& Schlichenmaier (2013) on Hinode degraded Stokes V profiles to retrieve opposite polarity fields.
The method considers asymmetric 3-lobe Stokes V profiles combined with Hinode SP far wing magnetograms. Since the method consider 3-lobee Stokes V
profiles in the redwing thus, these profiles preferentially located at downflow regions with opposite polarity (see Figs. 1 \& 3 of Franz \& Schlichenmaier (2013))

The results presented in the following section are for
synthetic line profiles calculated along vertical lines of sight,
corresponding to observations exactly at disk center such that they can be
compared with observations (Franz \& Schlichenmaier 2009, 2013) at disc center. We
discuss the analysis of the simulation with 32 km horizontal grid spacing in detail and compare results
briefly with those obtained from the simulation with 12 km grid spacing.

\begin{figure*}
\vspace{-130mm}
\centering
\includegraphics[width=240mm,angle=0]{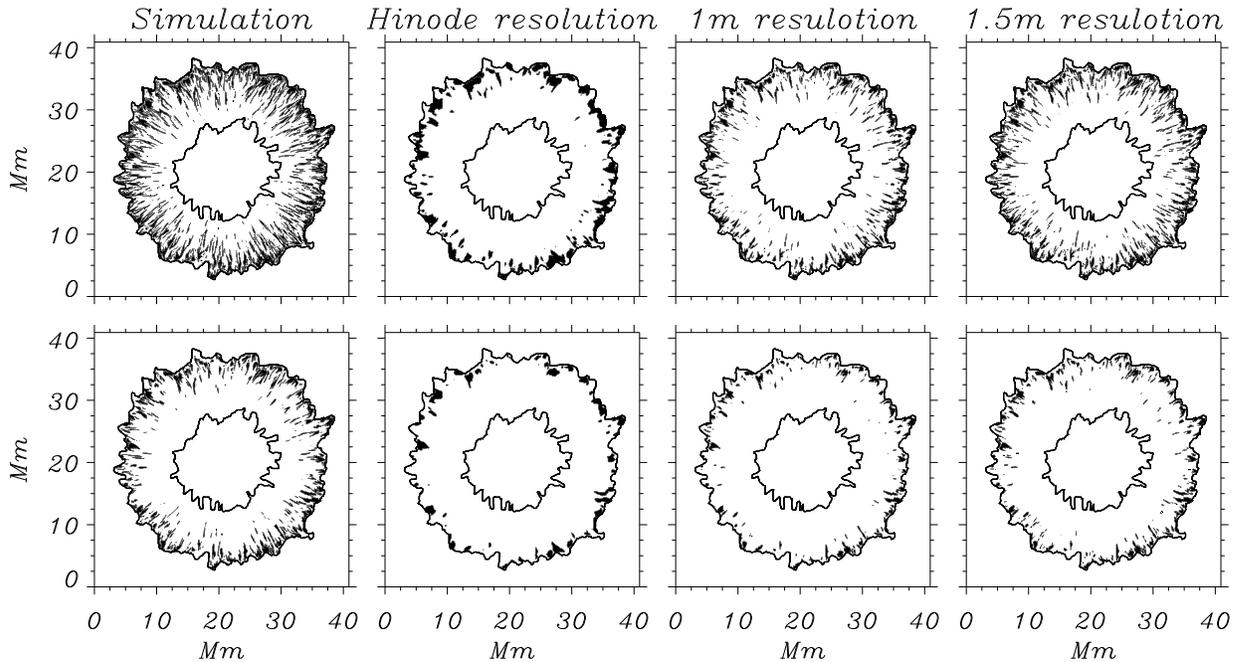}
\vspace{-15mm}
\caption{Opposite polarity field in simulated penumbra at $\tau=1$ (top row) and $\tau=0.1$ level. First column: maps at native resolution in the simulations. Second, third and fourth column: maps at Hinode (0.5 m), 1 m and 1.5 m resolution respectively.}
\end{figure*}

\begin{table*}
  \caption{Fraction of inverse polarity and downflow for 32 km and 12 km run}
  \begin{center}
    \begin{tabular}{ l  c c c c }
      \hline
      \hline
                     &     & 32 km              &                & 12 km \\
      \hline
      Resolution  & $\tau=1$ & $\tau=0.1$ & $\tau=1$ & $\tau=0.1$\\
      \hline
      Opposite polarity (at native resolution) & 21$\%$ &   11$\%$  & 20$\%$ &   12$\%$\\
      Opposite polarity (at Hinode (0.5 m) resolution) & 11$\%$ &   5$\%$   & 11$\%$ &   6$\%$\\
      Opposite polarity (at 1 m resolution) & 11$\%$ &   5$\%$      & 10$\%$ &   5$\%$\\
      Opposite polarity (at 1.5 m resolution) & 14$\%$ &   6$\%$   & 13$\%$ &   7$\%$\\
      Downflows (at native resolution) & 43$\%$ &   34$\%$          & 40$\%$ &   33$\%$\\
      Downflows (at Hinode (0.5 m) resolution) & 32$\%$ &   25$\%$          & 31$\%$ &   25$\%$\\
      Downflows (at 1 m resolution) & 45$\%$ &   37$\%$             & 41$\%$ &   34$\%$\\
      Downflows (at 1.5 resolution) & 46$\%$ &   37$\%$          & 42$\%$ &   35$\%$\\
      Downflows/Opposite polarity (at native resolution) & 70$\%$ &   49$\%$  & 71$\%$ &   49$\%$\\
      Downflows/Opposite polarity (at Hinode (0.5 m) resolution) & 88$\%$ &   73$\%$  & 91$\%$ &   79$\%$\\
      Downflows/Opposite polarity (at 1 m resolution) & 87$\%$ &   71$\%$     & 92$\%$ &   77$\%$\\
      Downflows/Opposite polarity (at 1.5 m resolution) & 83$\%$ &   64$\%$  & 87$\%$ &   68$\%$\\
      \hline
      \hline
      \label{tab:t2}
    \end{tabular}
  \end{center}
\end{table*}

\begin{figure*}
\vspace{-3mm}
\centering
\includegraphics[width=160mm,angle=0]{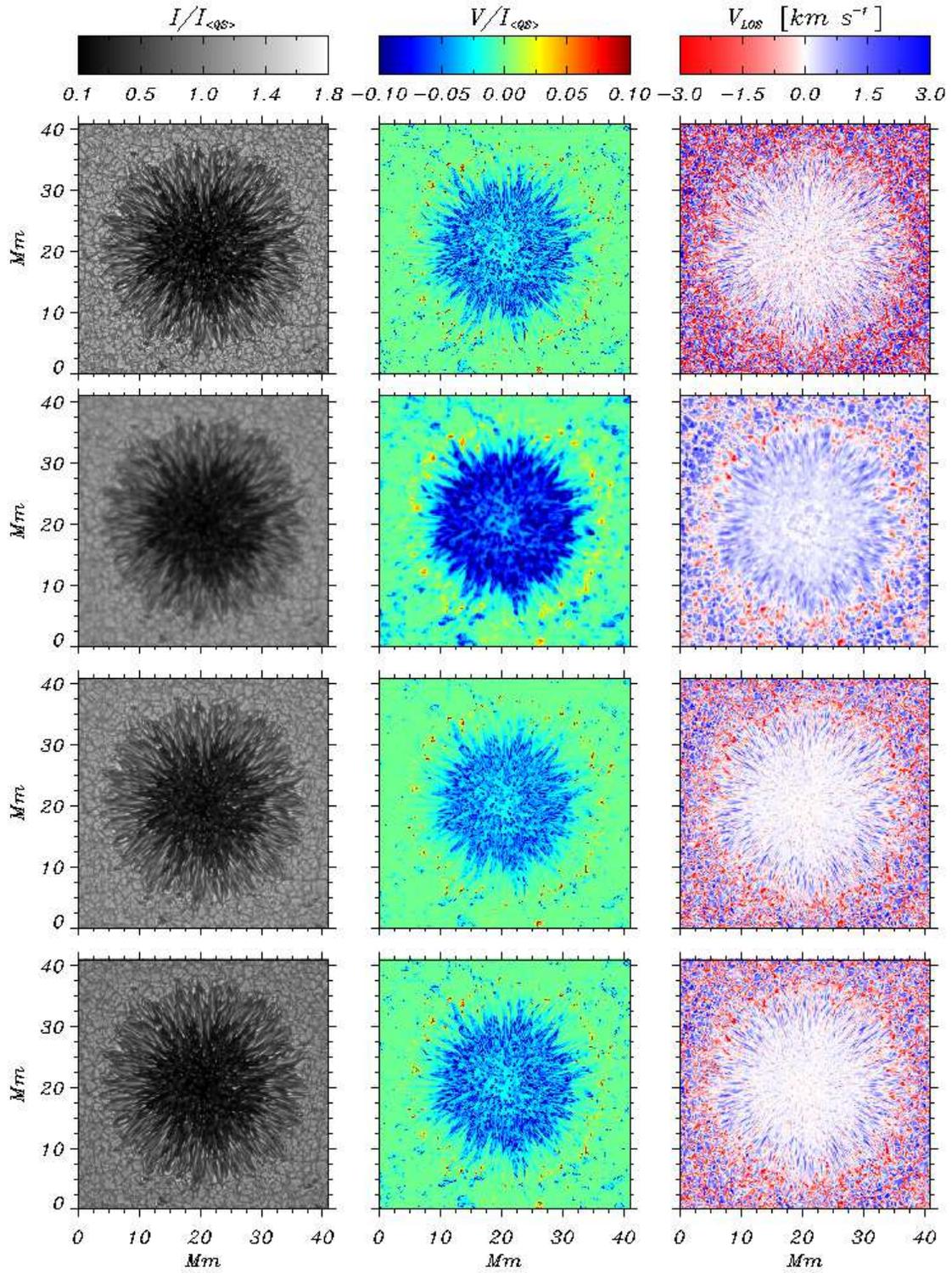}
%\vspace{-7mm}
\caption{Left panels: Normalized continuum intensity at 630 nm. Middle panels: Stokes V far wing magnetogram. Right panels: bisector velocity at 80$\%$ for FeI 6301.5 \AA~. First row: at simulation resolution. Second, third and fourth row: at Hinode (0.5 m), 1 m and 1.5 m resolution.}
\end{figure*}

\begin{figure*}
\vspace{-0mm}
\centering
\includegraphics[width=160mm,angle=0]{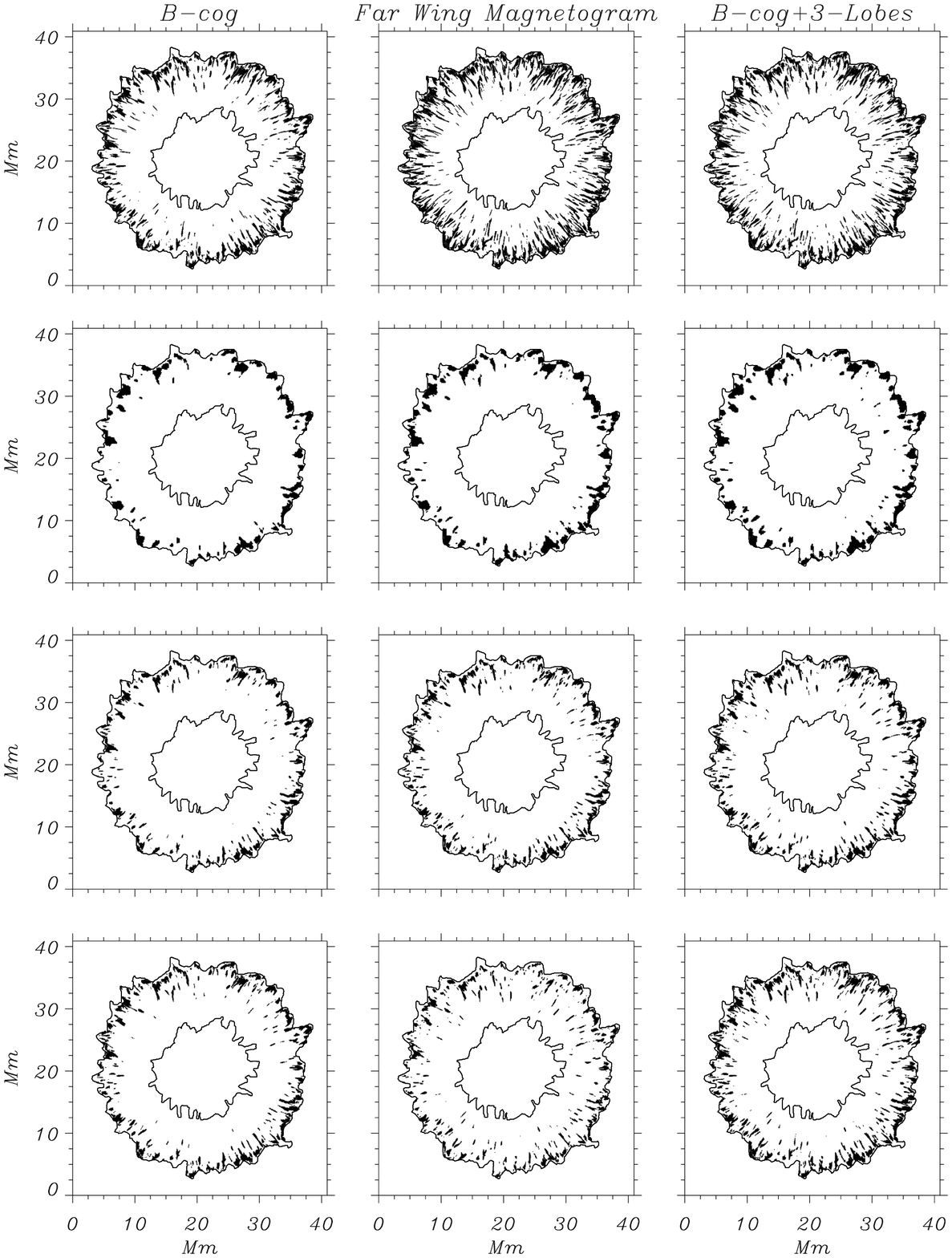}
\vspace{-7mm}
\caption{Left panels: opposite polarity maps from center-of-gravity method. Middle panels: maps far wing magnetogram. Right panels: maps for center-of-gravity \& 3-lobes. First row: at native resolution. Second, third and fourth row: at Hinode (0.5 m), 1 m and 1.5 m resolution.}
\end{figure*}

\section{Analysis}

We construct binary masks in order to retrieve the filling factors of opposite polarity field at simulation,
Hinode (0.5 m), 1 m and 1.5 m resolution. The inner and the outer boundary of masks are determined
by taking into account both the intensity and $B_{\rm LOS}$ (at $\tau=1$). The mask found to be well consistent with synthetic magnetogram and intensity without smearing. The masks exclude sunspot umbra and granulation beyond outer penumbra thus only contains penumbral part of the sunspot. We apply the above described 4 methods (bisector, center of gravity, wing magnetogram, 3-lobe) to synthetic observations at three telescope resolutions. We compare the retrieved velocity field and magnetic field to the simulation data in order to quantify how well a particular method can retrieve these quantities.

\subsection{Analysis based on data extrated on tau surfaces}

 We degraded line-of-sight magnetic field maps at $\tau=1$ and $\tau=0.1$ for Hinode (0.5 m), 1 m and 1.5 m resolution. The opposite
polarity field mask for the simulation (at 32 km grid spacing) and degraded maps for Hinode (0.5 m) and 1 m
resolution is depicted in Fig. 1. The top and bottom rows display opposite polarity field at $\tau=1$ and $\tau=0.1$ respectively. The opposite polarity field at $\tau=1$ within penumbra is 21$\%$ of the total penumbral area while for Hinode (0.5 m) and 1 m it is 11$\%$. At $\tau=0.1$ this area is half at native, Hinode (0.5 m) and 1 m resolution. The opposite polarity fraction significantly increase for 1.5 m at $\tau=1$. The downflow area covered in penumbra is 43$\%$, 32$\%$, 45$\%$ and 46$\%$ respectively. At $\tau=0.1$ the downflow area is lower. The fraction of opposite polarity pixels that are found in downflow regions is 70$\%$, 88$\%$, 87$\%$ and 83$\%$ at $\tau=1$ for native, Hinode (0.5 m), 1 m and 1.5 m resolution. This fraction is lower at $\tau=0.1$.  These quantities are tabulated in Table 1. Due to spatial smearing opposite
polarity field patches are only visible in the middle and outer penumbra at Hinode (0.5 m) resolution while
these patches are more clearly visible in the inner
(very few), middle and outer penumbra at the better 1 m and 1,5 m resolution at both $\tau$ levels. These quantities remains robust with grid spacing (12 km) as tabulated in the Table 1. However, spatial smearing leads to an overestimation of the association of opposite polarity magnetic field patches with downflows. Our findings are in agreement with Joshi et al. (2017) (cf. Figs. 22 and 23) where the authors find similar results at different optical depths in a simulated sunspot and in a sunspot observed from Hinode (0.5 m). Since Joshi et al. (2017) employs the spatially coupled inversion technique of van Noort (2012), they were able to detect opposite polarity field in the penumbra already at Hinode (0.5 m) resolution. However, without special processing observations show  opposite polarity field at Hinode (0.5 m) resolution only in the outer penumbra (Feanz \& Schlichenmaier 2009, 2013). This is consistent with our analysis, in which we do not deconvolve the synthetic observations.

\subsection{Analysis based on far-wing magnetograms and LOS velocity from bisector}

The left panels of Fig. 2 show the simulated sunspot (with 32 km grid spacing) in the continuum for the Fe\sci 6301.5 \AA~ line. The intensities
are normalized with respect to the average intensity of the plage area surrounding the sunspot.
The panels in the middle and right-handed columns of Fig.2 show the LOS velocity at 80$\%$ bisector level and Stokes V magnetogram, respectively.
The top row shows the results at the original resolution of the
simulation (32 km grid spacing). While only a few opposite polarity patches are visible
in the inner and middle penumbra, a significant area is covered in the outer penumbra.
Convective downflows along the sides of the penumbral filaments can be seen from the inner to the
outer penumbra. In outer penumbra downflows form larger patches mostly located at the end of filaments.

The second row of Fig. 2 shows these quantities after smearing to Hinode (0.5 m) resolution. Opposite polarities in the degraded Stokes V
magnetograms are visible in the middle and outer penumbra. Due to spatial
smearing the size of opposite polarity patches seen is larger compared to the
size at original simulation. Most of the downflow patches at the sides of the filaments are
not visible in the LOS map in the inner penumbra. Some downflow patches in the middle penumbra can be
seen with low amplitude. In the outer penumbra downflow patches show a larger area due to spatial
smearing.

The visibility of opposite polarity field and downflows is much better
for higher 1 m and 1.5 m resolution as illustrated in the third and fourth row of Fig. 2. The opposite polarity field at the sides of
a few filaments is visible in the inner penumbra. In the middle penumbra more elongated patches can be seen. In the outer penumbra
more opposite polarity field with a sharp boundary compared to the Hinode (0.5 m) map is visible. The elongated downflows at the sides of filaments can be seen in the whole penumbra. This is in agreement with Tiwari et al. (2015) where the spatially coupled inversion technique
of van Noort (2012) significantly improves the resolution, such that opposite polarity field and downflow patches at the sides of penumbral filaments can be seen already in the whole penumbra. Without such processing a higher resolution, corresponding to about a 1 m telescope is required.

%\begin{figure*}
%\vspace{-10mm}
%%\hspace{9mm}
%\centering
%\includegraphics[width=120mm,angle=0]{figure2.eps}
%\vspace{-7mm}
%\caption{Masks used for the opposite polarity field derived from the Stokes V magnetograms. Left: simulation; Middle: Hinode; Right: SST.}
%\end{figure*}

      The opposite polarity field occupies 17$\%$ of the total penumbral pixels in the synthetic Stokes V
      magnetogram at native resolution. Degraded Hinode (0.5 m), 1 m and 1.5 m covers 8$\%$, 6$\%$ and
      7$\%$ of the penumbral area, respectively.

      The downflow fraction is 42 $\%$ in the synthetic LOS velocity at native resolution. Degraded Hinode (0.5 m), 1 m and 1.5 m
      occupy 14$\%$, 35$\%$and 38$\%$, respectively.

      We also calculated the association of opposite polarity field with downflows.
      The fraction of opposite polarity fields that harbor downflows are 67$\%$ in the synthetic maps at native resolution. This fraction is
      72$\%$, 92$\%$ and 90$\%$ respectively for Hinode (0.5 m), 1 m and 1.5 m degraded maps.

      We repeat the same analysis for the simulation with 12 km horizontal and 8 km vertical grid spacing and find that with increasing the grid spacing in simulation the amount of opposite polarity, downflow and fraction of downflow that associated with opposite polarity do not change much as suggested by Rempel (2012). The comparison is
      illustrated in Table 2. The fraction of opposite polarity field present in the
      synthetic magnetograms at native resolution is almost the same for both runs. It
      increases from  8 to 10$\%$, 6 to 12$\%$ and from 7 $\%$ to 8$\%$ in the Hinode (0.5 m), 1 m and 1.5 m degraded maps, respectively.
      The downflow fraction at native resolution is
      reduced from 42 to 40$\%$ in the synthetic map at 12 km resolution. The downflow fraction increased
      from 14 to 23$\%$ in degraded Hinode maps. In the 1 m and 1.5 m degraded maps this fraction remains almost the same.
      The fraction of opposite polarity that is associated with downflow
      increases from 67 to 69$\%$ for the 12 km resolution synthetic map at native resolution.
      This fraction is increased from 72 to 85$\%$, reduced from 92 to 83$\%$ and remains same in the Hinode (0.5 m), 1 m and 1.5 m degraded maps, respectively. Overall these quantities are robust
      in both the 32 km and 12 km run, most of the differences occur due to spatial smearing and noise.

      We also tested the effect of noise on these quantities and noticed that a significant amount of small scale magnetic and velocity field is elusive due to typical instrumental noise criteria. However, it is actually present in the simulated data. These quantities are also tabulated in the Table 2 and show higher values without noise seclection.

      \begin{table*}
  \caption{Fraction of inverse polarity and downflow with as well as without noise for farwing magnetograms and higher bisectors}
  \begin{center}
    \begin{tabular}{ l  c c c c}
      \hline
      \hline
                  &        & With noise                    &          & Without noise \\
      \hline
      Resolution  & 32 km & 12 km & 32 km & 12 km \\
      \hline
      Opposite polarity (native resolution) & 17$\%$ & 17$\%$ & 17$\%$ & 17$\%$\\
      Opposite polarity (at Hinode (0.5 m) resolution) & 8$\%$ & 10$\%$ & 14$\%$ & 17$\%$\\
      Opposite polarity (at 1 m resolution) & 6$\%$ & 12$\%$ & 10$\%$ & 12$\%$\\
      Opposite polarity (at 1.5 m resolution) & 7$\%$ &   8$\%$   & 12$\%$ &   13$\%$\\
      Downflows (native resolution) & 42$\%$ & 40$\%$ & 49$\%$ & 46$\%$\\
      Downflows (at Hinode (0.5 m) resolution) & 14$\%$ & 23$\%$ & 23$\%$ & 37$\%$\\
      Downflows (at 1 m resolution) & 35$\%$ & 36$\%$ & 44$\%$ & 45$\%$\\
      Downflows (at 1.5 m resolution) & 38$\%$ &   38$\%$          & 46$\%$ &   46$\%$\\
      Downflows/Opposite polarity (native resolution) & 67$\%$ & 69$\%$ & 71$\%$ & 73$\%$\\
      Downflows/Opposite polarity (at Hinode (0.5 m) resolution) & 72$\%$ & 85$\%$ & 67$\%$ & 78$\%$\\
      Downflows/Opposite polarity (at 1 m resolution) & 92$\%$ & 83$\%$ & 85$\%$ & 87$\%$\\
      Downflows/Opposite polarity (at 1.5 m resolution) & 90$\%$ &   90$\%$  & 81$\%$ &   82$\%$\\
      \hline
      \hline
      \label{tab:t1}
    \end{tabular}
  \end{center}
\end{table*}

  \subsection{Analysis based COG and COG-3-Lobes methods}

      Schlichenmaier \& Franz (2013) showed that asymmetric 3-lobe Stokes V profiles
      combined with Hinode SP magnetograms can retrieve opposite polarity magnetic field
      efficiently. We applied the algorithm used by Schlichenmaier \& Franz (2013) on synthetic maps at native resolution, Hinode (0.5 m), 1 m and 1.5 m resolution. Center-of-gravity, far wing magnetogram and center-of-gravity+3-Lobes maps (from left to right columns) for all four cases (top to bottom rows) are displayed in Figure 3. The opposite polarity field for center-of-gravity+3-Lobes maps is higher compared to far wing magnetograms for Hinode (0.5 m), 1 m and 1.5 m maps (see Table 3) which is consistent with Schlichenmaier \& Franz (2013). At native resolution both maps are comparable. Comparing with Figure 1 and Table 3 we find that the fraction of opposite polarity field is comparable in maps at $\tau=1$ for Hinode (0.5 m), 1 m and 1.5 m cases. According to Franz \& Schlichenmaier (2009, 2013) the opposite polarity field is underestimated by the methods used in the observational analysis. The fraction of opposite polarity in downflows is lower in our analysis. This could be due to the fact that the amount of opposite polarity field increase by chosen scheme but detected downflow remain same. Possibly these quantities are measured at different level in the atmosphere as the far wing and higher line bisector are located in deep photosphere but slightly above $\tau=1$ level. Small scale magnetic field can't be detected due to spatial smearing which gives lower fraction of opposite polarity magnetic field.

\begin{table}
  \caption{Fraction of inverse polarity and downflow for COG and 3-lobe profiles }
  \begin{center}
    \begin{tabular}{ l  c  }
      \hline

               Resolution &   COG and 3-lobe profiles \\
      \hline
      Opposite polarity (at native resolution) & 19$\%$\\
      Opposite polarity (at Hinode (0.5 m) resolution) & 10$\%$\\
      Opposite polarity (at 1 m resolution) & 8$\%$\\
      Opposite polarity (at 1.5 m resolution) & 9$\%$\\
      Downflows/Opposite polarity (at native resolution) & 63$\%$\\
      Downflows/Opposite polarity (at Hinode (0.5 m) resolution) & 69$\%$\\
      Downflows/Opposite polarity (at 1 m resolution) & 81$\%$\\
      Downflows/Opposite polarity (at 1.5 m resolution) & 83$\%$\\
      \hline
      \hline
      \label{tab:t3}
    \end{tabular}
  \end{center}
\end{table}

The amount of opposite polarity magnetic field and downflows in our degraded
  maps are likely overestimated since we consider ideal instrument and seeing conditions.
  Ground based observations are used after post reconstruction
  techniques and stray light removal (cf. Joshi et al. 2011, Scharmer et al. 2011, 2013, Scharmer \& Henriques 2012).
  Schlichenmaier \& Franz (2013) raised concerns that stray light removal from a 2D
  spectrograph can affect the detection of downflows in the penumbra. However, Scharmer (2014) validated the stray light removal method used for 1 m observations. Scharmer et al. 2013 find a large number of elongated opposite polarity field patches in the inner penumbra compared to the simulations presented here (see Fig. 11 of Scharmer et al. 2013).

\subsection{Individual case study}

  \begin{figure*}
\vspace{-4mm}
%\hspace{9mm}
\centering
\includegraphics[width=160mm,angle=0]{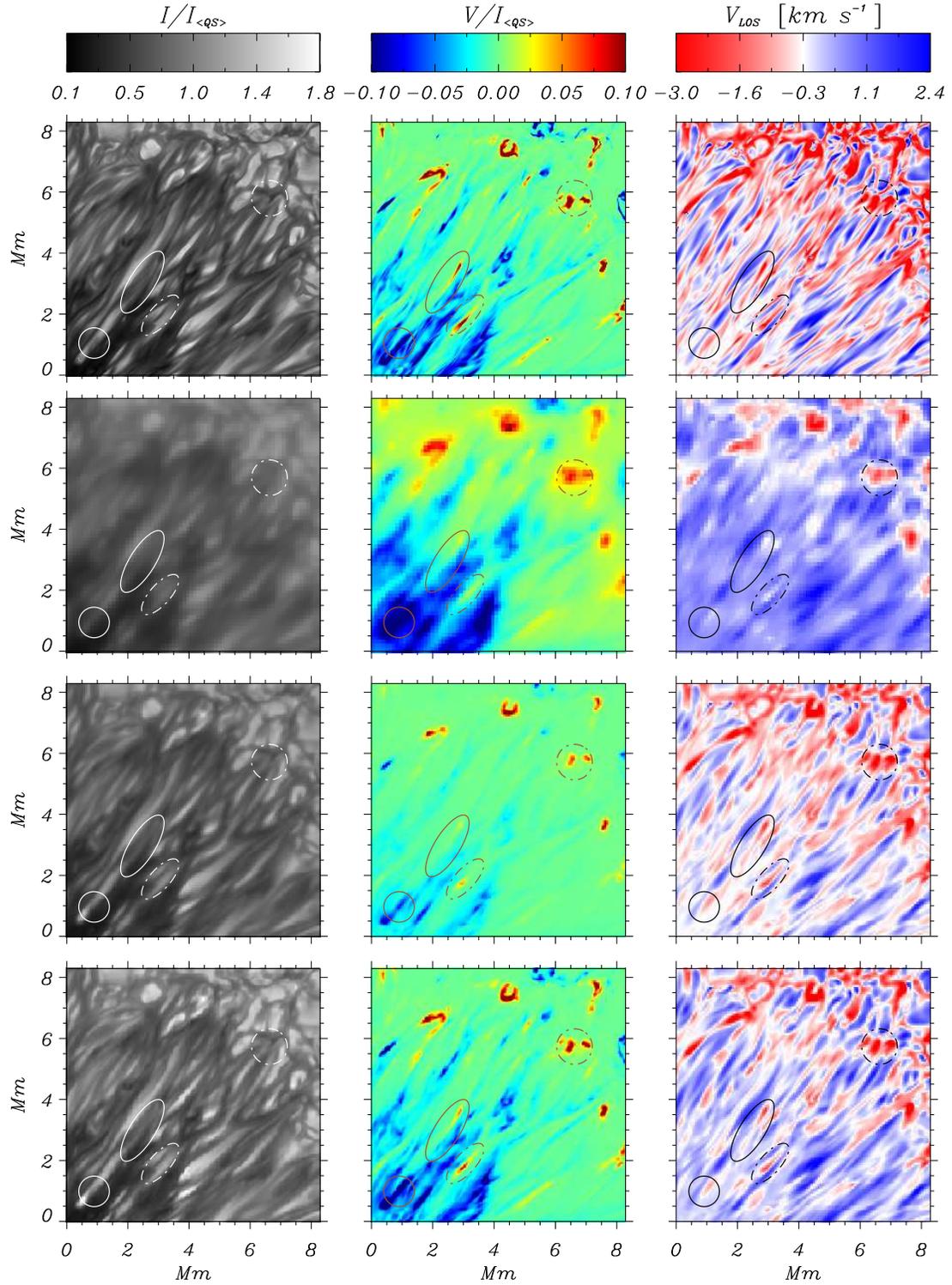}
\vspace{-7mm}
\caption{Selected section of penumbra. Left panel: normalized intensity.
Middle panel: far wing magnetogram. Right panel: Line of sight velocity.
Different locations in selected filaments are indicated by a solid and dashed circle and an ellipse. First row: simulation resolution. Second to fourth row: Hinode (0.5 m), 1 m and 1.5 m resolution.}
\end{figure*}

   In this section we discuss in detail selected  opposite polarity patches and downflows found at
    different locations in the simulated penumbra.

    Figure 4 shows a section of penumbra from the 32 km run. The left panel depicts normalized intensity, while the
    middle and right panels illustrate far wing magnetograms and LOS
    velocity, respectively. The first row shows quantities at the simulation resolution, while the second, third and fourth rows show the same
    quantities for maps degraded to Hinode (0.5 m), 1 m and 1.5 m resolution, respectively. A typical simulated penumbral filament in intensity shows higher brightness in the inner part (head), which decreases towards the middle and the outer part (tail). The Stokes V signal is low in the inner part (head) and increases towards the outer part (tail). Upflows along the axis of a filament are higher at the inner part (head) and reduce towards the outer part (tail). The uplows is followed by downflows along the lateral edges of filaments (Rempel et al. 2009 a, b). At the outer part (tail) of a filament strong downflows drag magnetic field lines downwards and magnetic field reversal appears (Rempel 2011). These characteristics are consistent with observations in intensity (Bharti et al. 2010, Tiwari et al. 2013), magnetic field and velocity (Tiwari et al. 2013) of penumbral filaments. From these properties, different parts of a penumbral filaments can be recognized.  A solid circle and ellipse indicate the  location of the inner and outer part of a filament. We can see that there is no signal of opposite polarity in the inner part of the filament (solid circle), while downflows are clearly visible at the 1 m and 1.5 m resolution.
    Downflows and opposite polarity at the outer part of this filament (solid ellipse) are
    visible at Hinode  (0.5 m), 1 m and 1.5 m resolution. Similarly downflows and opposite
    polarity field in the middle part of a filament (indicated by dashed ellipse)
    is visible in all maps. An example of a downflow near the outer edge of the
    penumbra is indicated by a dashed circle. This
    downflow is associated with opposite polarity field and is clearly visible in
    all maps. It is to be noted that spatial smearing
    increases the area of downflow patches, particularly in the outer part of
    the penumbra. This is the reason why we see discrepancies in the quantities
    shown in Table 2.

\section{Discussion and conclusion}

In this work, we used spatialy and spectraly degraded Synthetic Stokes profiles according to telescope resolution and instrumental properties. Different methodes used on these profiles to retrieve opposite polarity and downflow present in the sunspot penumbra to find dependence on resolution and method applied. The amount of opposite polarity magnetic field and downflow strongly depends on telescope resolution and method adopted. The main results are :

\begin{enumerate}

\item We confirmed the findings of Scharmer \& Henriques (2012) that convective
downflows can be seen clearly in FeI 6301.5 \AA~ line at SST (1 m) resolution.
The fraction of opposite polarity field present in the simulated penumbra and their
 association with downflows is in agreement with Ichimoto at al (2007) and
 Franz \& Schlichenmaier (2013)  for the Hinode (0.5 m) case.

\item The appearance of opposite
 polarity field and downflows are patchy rather than elongated,
which is in agreement with Franz \& Schlichenmaier (2013) for Hinode (0.5 m) degraded maps.
However, these patches are radially elongated along filament's sides in 1 m degraded maps
as shown by Scharmer et al. (2013). This different appearance in observations is
due to spatial smearing. Ruiz Cobo \& Asensio Ramos 2013 deconvolved spectropolarimetric
observations of Hinode (0.5 m) using a new regularized method and found that most of the penumbral
filaments have opposite polarity field at their sides in the whole penumbra. Tiwari et al. 2013
used spatially coupled 2D inversion scheme (van Noort 2012) on Hinode spectropolarimetric
data and reported a similar result. Using the same inversion scheme van Noort (2013) reported
 very high downflow velocities and very high magnetic field in the outer part of the
 penumbra. The presence of opposite polarity field and associated downflows is
confirmed using these new inversion schemes. The absence of downflows at the edge of penumbral filaments
and opposite polarity in the analysis of Franz \& Schlichenmaier (2009, 2013) in the inner and middle
penumbra is due to spatial smearing. Properly retrieving downflows and opposite polarity field at Hinode (0.5 m)
resolution requires more advanced inversion techniques.

\item The amount of opposite polarity magnetic field and downflows are comparable for 32 km and 12 km grid resolution in simulations, which suggests the robustness of these quantities in the simulated sunspot penumbra. This confirms the findings of Rempel (2012) where author
    compares physical parameters in detail at 32 km and 12 km grid resolution.

\item The fraction of opposite polarity magnetic field depends also on the choice of Stokes profiles. Franz \& Schlichenmaier (2013) included 3-lobe Stokes profiles, which increases the detection of the fraction of opposite polarity field in the penumbra. Here we find that 3-lobes method significantly detects opposite polarity field associated with convective downflows which are in agreement with  Franz \& Schlichenmaier (2013).

\item The methods used in observations are based on assumptions that profiles are associated with strong gradients in both the line-of-sight magnetic field and the velocity such that they correspond to strong downflows and opposite polarity magnetic fields in the deep photosphere. Far wing magnetogram and 3-lobes method (Franz \& Schlichenmaier (2009, 2013)) thus selectively will lead to the identification of patches in the penumbra where opposite polarity magnetic field is associated with downflows. The observational data possibly affected by blends at far wing and at higher bisector level (Esteban et al. 2015). The fraction of opposite polarity at $\tau=1$ serves as the reference and suggests that it depends on methods used in observations and depends on resolution of the telescope. The small-scale field in inner and middle penumbra is not detected by these methods due to spatial smearing.

\end{enumerate}

The association of opposite polarity magnetic field and downflows is an essential part of
overturning convection. Strong downflows at the sides of the filaments
bend magnetic field lines downward and thus opposite polarity magnetic field appears at the sides of the
filaments (c.f. figure 22 of Rempel 2012). Bharti et al. (2010) first reported
a similar mechanism for opposite polarity magnetic field at the sides of umbral dots. In the outer
part of filaments the strong horizontal Evershed flow component returns beneath the photosphere
in more concentrated downflow patches that show a significant amount of opposite polarity magnetic field (Rempel 2012, van Noort 2013, Esteban et al. 2015).

Using infrared lines at GREGORE (Schmidt et al. 2012) 1.5 meter telescope Franz et al. (2016) found that Stokes V profiles associated with opposite polarity are smaller by a factor of two compared to visible. The authors suggested it is due to the fact that strong flux concentrations at outer penumbra associated with downflows cause evacuation and spectral lines formed deeper inside such flux concentrations are narrower with depth. Thus, the area of flux concentrations is lower for deeply forming spectral lines. This difference also occurs due to larger height coverage by visible lines while infrared lines form in the narrow region. This hypothesis can be tested with synthetic infrared lines in future work. We are aware of the fact that quantitative difference for different method occur due to formation height of spectral lines and detection of these quantities at different surface in the photosphere.

Our investigation highlights that small-scale magnetic structure are strongly influenced by resolution and instrumental noise of observations. A proper quantification of opposite polarity magnetic field and resolution with convective downflows will require high polarimetric sensitivity and spatial resolution such as produced by DKIST in the future. In order to extend this study to DKIST we will require higher resolution simulations.

\acknowledgments
 We thanks Dr. Michiel van Noort for providing us Hinode PSF
and Dr. Franz Morten for 3-lobe finding algorithm.
The authors thank Dr. R. Centeno-Elliot and Andreas Lagg for helpful comments on
the manuscript and anonymous referee for constructive comments to improve the presentation
of the manuscript. LB is grateful to HAO visitor program for supporting his visit.
The National Center for Atmospheric Research (NCAR) is sponsored by the
National Science Foundation. Computing resources for the sunspot model
utilized in this investigation were provided through NCAR's Computational
Information Systems Laboratory (CISL) as well as through the National Science
Foundation (NSF) at the National Institute for Computational Sciences (NICS) under
grant TG-AST100005.

\end{document}